\documentclass[a4paper,11pt]{article}
\usepackage{jinstpub}
\pdfoutput=1

\title{\boldmath  A semi-analytical energy response model for low-energy events in JUNO}

\usepackage{amsmath}
\usepackage{subfiles}

\newcommand{\notinsubfile}[1]{}
\newcommand{\printbibliography}{
\bibliographystyle{jinst}
\bibliography{References/ref,References/ref_gna}
}

\usepackage{graphicx}
\usepackage{rotating}
\usepackage{hyperref}
\usepackage[latin1]{inputenc}
\usepackage{mathtools}
\usepackage{color}
\usepackage{lineno}
\usepackage{url}

\def\D{\mathrm{d}}

\author[a,b,1]{P. Kampmann, \note{Corresponding author.}}
\author[a,c,2]{Y. Cheng, and \note{Beijing Institute of Spacecraft Environment Engineering, Beijing 100094, China}}
\author[a,b]{L. Ludhova}

\affiliation[a]{Institut f\"ur Kernphysik, Forschungszentrum J\"ulich, 52425 J\"ulich, Germany}
\affiliation[b]{RWTH Aachen University, 52062 Aachen, Germany}
\affiliation[c]{Institute of High Energy Physics, 100094 Beijing, China}

\emailAdd{p.kampmann@fz-juelich.de}

\abstract {The Jiangmen Underground Neutrino Observatory (JUNO) is a next-generation neutrino experiment under construction in China expected to be completed in 2022. As the main goal it aims to determine the neutrino mass ordering with 3-4\,$\sigma$ significance using a 20\,kton liquid scintillator detector. It will measure the oscillated energy spectrum of electron anti-neutrinos from two nuclear power plants at about 53\,km baseline with an unprecedented energy resolution of 3\% at 1\,MeV. A requirement of the JUNO experiment is the knowledge of the energy non-linearity of the detector with a sub-percent precision. As the light yield of the liquid scintillator is not fully linear to the energy of the detected particle and dependent on the particle type, a model for this light yield is presented in this paper. Based on an energy non-linearity model of electrons, this article provides the conversion to the more complex energy response of positrons and gammas. This conversion uses a fast and simple algorithm to calculate the spectrum of secondary electrons generated by a gamma, which is introduced here and made open access to potential users. It is also discussed how the positron non-linearity can be obtained from the detector calibration with gamma sources using the results presented in this article.}

\keywords{Neutrino detectors; Detector modelling and simulations I; Detector modelling and simulations II; Ionization and excitation processes; Liquid detectors; Photoemission; Scintillators, scintillation and light emission processes; Simulation methods and programs;  }
\arxivnumber{xxxx}

\begin{document}
\maketitle

\section{Introduction}
\label{sec:Intro}

The Jiangmen Underground Neutrino Observatory (JUNO) is a neutrino experiment under construction in China expected to be completed in 2022~\cite{yellowbook}.
As a next generation neutrino experiment it aims to address current challenges in neutrino and astroparticle physics. The detector, a 20\,kt liquid scintillator tank instrumented with 18,000 photomultiplier tubes (PMTs), will be placed at a distance of 53\,km from the Taishan and the Yangjiang nuclear power plants. It will be placed underground with an overburden of about 700\,m rock.

The 53\,km distance optimizes the sensitivity to the neutrino Mass Ordering (MO) determination, which is the main goal. JUNO aims to address it with 3-4\,$\sigma$ significance. In addition to that, JUNO aims to measure the oscillation parameters $\theta_{12}$, $\Delta m^2_{21}$, and $\lvert\Delta m^2_{32}\rvert$ with a sub-percent precision. These measurements rely on measuring the coincident signal from Inverse Beta Decay (IBD) events which consist of a prompt positron and a delayed gamma signal due to neutron capture.

Apart from measuring the neutrinos from the nuclear reactors, JUNO plans to measure the neutrinos from various other natural sources.
These include the measurement of geo-neutrinos, solar neutrinos, atmospheric neutrinos, supernova neutrinos, and the diffuse supernova neutrino background.
JUNO will also search for the proton decay.
The physics programme is described in detail in~\cite{yellowbook}.

To reach its goals, the JUNO experiment needs to fulfill several requirements. Among the most challenging ones are those on the energy scale and resolution. It is required to reach a relative uncertainty on the energy scale of less than 1\% and an energy resolution better than 3\% at 1\,MeV.
The light emission in a liquid scintillator is not linear with the deposited energy of particles and depends on the detected initial particle type.
As inappropriate modeling of the non-linearity biases the determination of the MO~\cite{yellowbook}, it is of imminent importance to address this topic carefully.
As the JUNO experiment will use gamma sources for the energy non-linearity calibration, it is important to develop methods how to derive non-linearity model for electrons and positrons based on the calibration data.

This article focuses on the description of the energy non-linearity in the scintillation medium.
It is a topic of general interest in organic scintillator physics with numerous publications on it~\cite{DYBCalib,DCNonLin,KamLANDnonlin,LABquenching}.
The non-linearity will be determined from the ratio of the visible energy over deposited energy.
The term \textit{visible energy} is used here to describe the expected amount of detectable light produced in the scintillation medium in the detector.

The article is organized as follows. Section~\ref{sec:juno} briefly reviews the physics motivation of the reactor anti-neutrino program (Sec.~\ref{sec:physics}) and the detector concept (Sec.~\ref{sec:detector}) of the JUNO experiment. In Section~\ref{sec:model} the modelling of the energy non-linearity effects is described. The part~\ref{sec:model_nl} deals with the light non-linearity in the measurement of electrons and of the kinetic energy loss for positrons (thus excluding the annihilation part) due to the ionisation quenching (Sec.~\ref{sec:quench}) and Cherenkov light production (Sec.~\ref{sec:Cher}). Since electrons directly produce scintillation photons through ionization, their model is simpler than for gammas or positrons, which also produce other secondary particles. 
The non-linearity model is based on the Birks' empirical formula~\cite{Birks}. Section~\ref{sec:model_gammasim} describes the conversion of the electron non-linearity model to the more complex non-linearity model for gammas and positrons including the annihilation. Gammas loose their energy in the scintillation medium under the production of several secondary electrons responsible for the emission of scintillation light.
An algorithm is presented to generate the energy distributions of the secondary electrons, as discussed in Sec.~\ref{sec:secondary_el}. These are then used to calculate the full non-linearity of the gammas by adding up the contributions from each secondary electron. This algorithm is validated and compared to the JUNO Geant4 simulation~\cite{Lin_2017} in Sec.~\ref{sec:gammasim_val}. It is shown, that it reproduces the results from the Geant4 simulation and has computational benefits due to its easy use and fast calculation (Sec.~\ref{sec:benfits}). Resulting non-linearity model for gammas is then presented in Sec.~\ref{sec:NL_gammas}. As positrons produce scintillation light directly in the deposition of their kinetic energy as well as through the annihilation producing two gammas as secondary particles, their non-linearity model can be constructed through the combination of these. This is shown in Sec.~\ref{sec:positron}.
In part~\ref{sec:posi_atrest}, the positron non-linearity is evaluated for the case of positron annihilation at rest into two photons. To evaluate the accuracy of this simplification, the resulting non-linearity is compared to the full JUNO Geant4 simulation with its more comprehensive physics description (Sec.~\ref{sec:posi_higherorders}). The final summary and outlook is then discussed in Sec.~\ref{sec:summary}.

\section{JUNO experiment}
\label{sec:juno}

\subsection{Physics goals with reactor anti-neutrinos}
\label{sec:physics}

 Nuclear reactors are a powerful source of electron anti-neutrinos ($\bar{\nu}_e$) with MeV energies. In several historical experiments they played a key role, that they preserve also today in the quest for answering open questions of neutrino physics. JUNO is the only experiment that aims to determine the yet unknown sign of $\Delta m^2_{32}$ mass splitting based on the oscillation of reactor anti-neutrinos in the vacuum dominated regime. We refer to this sign as the {\it Mass Ordering} (MO). The MO is of importance in the determination of the CP-violation phase $\delta$, of the Majorana phases in case the neutrino is a Majorana particle, and of the $\theta_{23}$-octant~\cite{NeutrinoStatus}.

\begin{figure}[t]
    \includegraphics[width=0.9\textwidth]{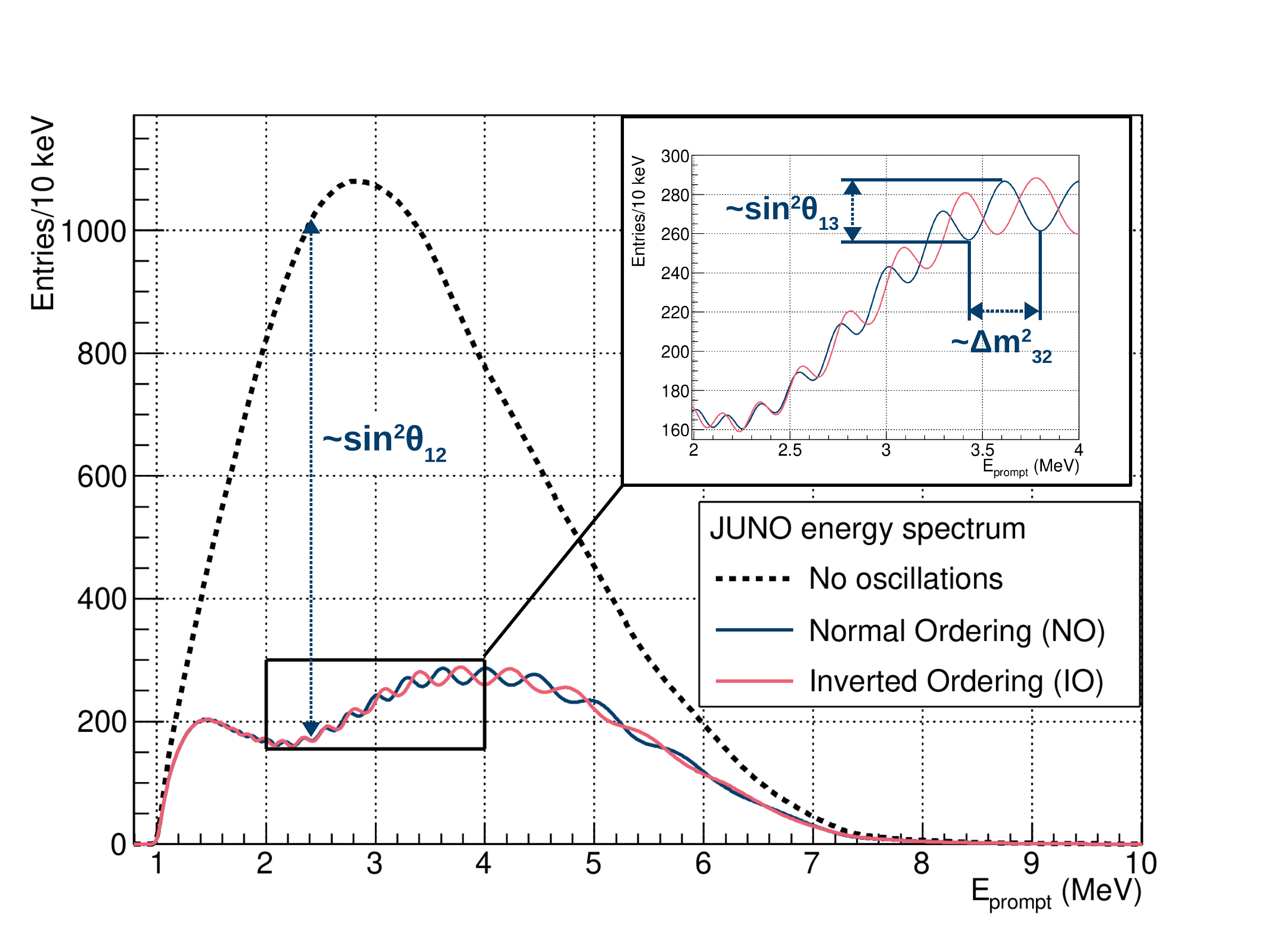}
    \caption{Expected energy spectra of the prompt IBD candidates in the JUNO experiment in no oscillation (dashed line) and three-neutrino oscillation models assuming normal (solid blue line) and inverted (solid red line) MO. The calculation was performed with the GNA software~\cite{Fatkina:2019ogr} assuming
    $3\%/\sqrt{E_\text{prompt}\text{[MeV]}}$ energy resolution and normal ordering oscillation parameters from \cite{PDG}. To display the inverted ordering, only the sign of $\Delta m^2_{32}$ was inverted. The insert shows in detail the energy region between 2 - 4\,MeV, which is the most dependent on the MO. The features of the spectra sensitive to $\theta_{12}$ and $\theta_{13}$ mixing angles are also demonstrated.
    }
	  \label{fig:junospectrum}
\end{figure}

To detect reactor anti-neutrinos, JUNO will use the Inverse Beta Decay (IBD) reaction on protons~\cite{IBD}:
\begin{equation}
  \bar{\nu}_e + p \rightarrow e^{+} + n
\label{eq:IBD}
\end{equation}
sensitive only to electron flavour and having a kinematic threshold of 1.806\,MeV. The cross section of the IBD interaction can be calculated precisely with an uncertainty of 0.4\%~\cite{strumia2003precise}. In this process, a positron and a neutron are emitted as reaction products. The positron promptly comes to rest and annihilates emitting two 511\,keV $\gamma$-rays, yielding a {\it prompt} signal. The neutron, after being thermalized, is captured ($\tau \sim$200\,$\mu$s) mostly on a proton, releasing a gamma with the constant energy of the deuteron binding energy of 2.2\,MeV, providing a {\it delayed} signal. A space and time coincidence between the prompt and delayed signals  significantly suppresses the background.

The JUNO sensitivity to MO 
is based on the dependence of the  oscillation pattern of reactor anti-neutrino energy spectrum on it. In the IBD interaction, the incident anti-neutrino energy $E_{\bar{\nu}_e}$ is directly correlated with a visible energy of the prompt signal $E_{\text{prompt}}$:
\begin{equation}
E_{\text{prompt}} \sim E_{\bar{\nu}_e}- 0.784\,\, \mathrm {MeV}.
\label{eq:Eprompt}
\end{equation}
This dependence is then exploited in the measurement of the $E_{\text{prompt}}$ spectrum, as depicted in Fig.~\ref{fig:junospectrum}. Consequently, the precision and accuracy of the reconstructed neutrino energy depends directly on the reconstruction of the positron energy. The distinction of the $E_{\text{prompt}}$ spectra corresponding to the normal and inverted MO allows its determination. Strict requirements on the energy reconstruction must be imposed to increase the significance of MO determination.
These include not only a resolution of less than 3\% at 1\,MeV, but also a sub-percent uncertainty on the energy scale.

 \subsection{The detector}
 \label{sec:detector}
 
\begin{figure}[!htb]
\includegraphics[trim=0 0 0 229,clip,width=0.9\linewidth]{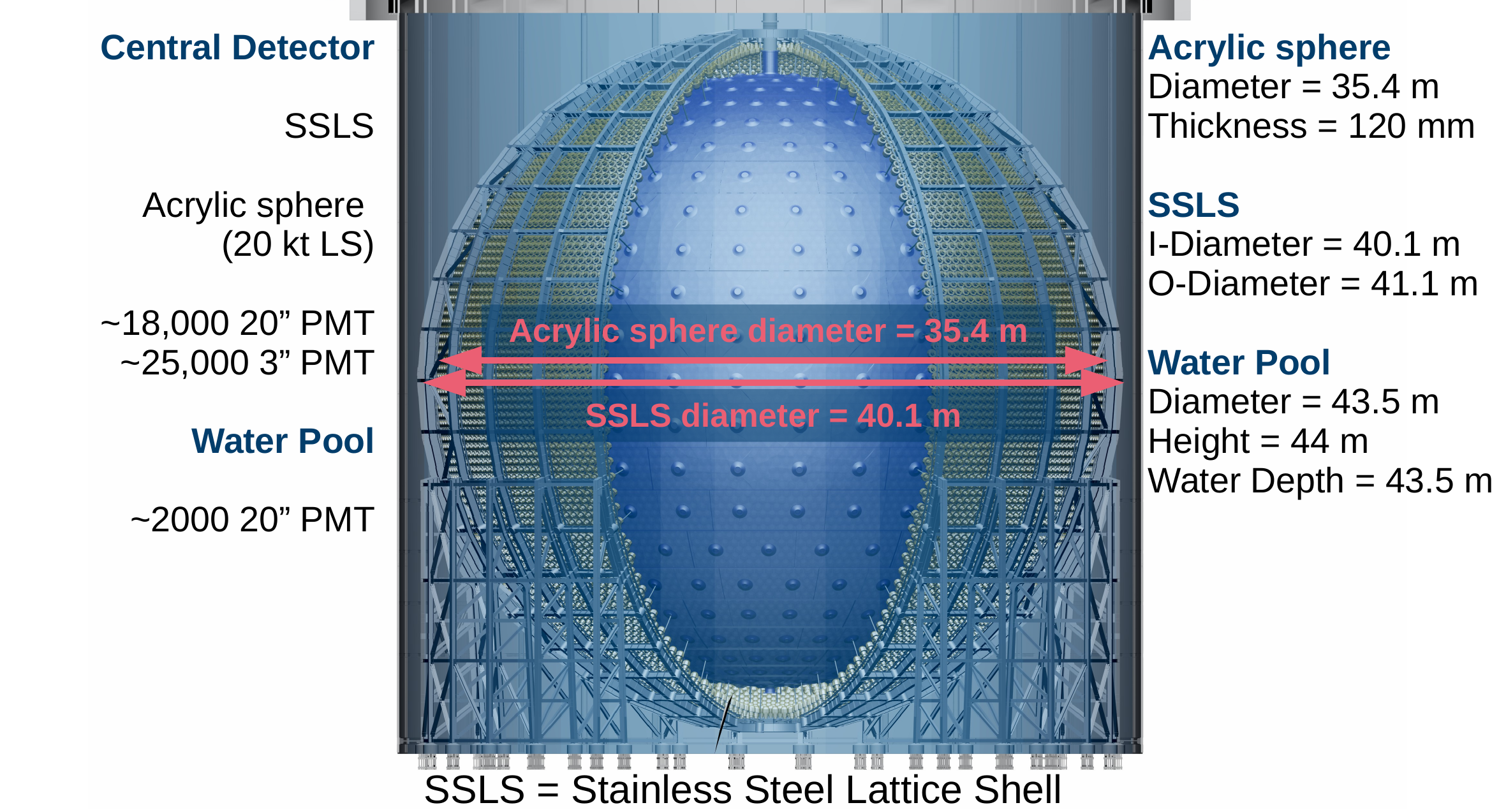}
\caption{The scheme of the JUNO detector.}
\label{fig:detector}
\end{figure}

The scheme of the JUNO detector setup is shown in Fig.~\ref{fig:detector}. The core {\it central detector} (CD) is composed of the {\it Acrylic Sphere} with 17.7\,m radius filled with 20\,kt of liquid organic scintillator (LS). The CD is equipped with about 18,000 large, 20'' and about 25,000 small, 3'' PMTs. 

The detection medium in the scintillator is linear alkylbenzene (LAB), a straight alkyl chain of 10-13 carbon atoms attached to a benzene ring~\cite{yellowbook}. As additives, 2.5\,g/l of PPO (2,5-diphenyloxazole) as fluor and 1-3\,mg/l bis-MSB as wavelength shifter will be used.
The density of the liquid scintillator mixture is expected to be 0.859\,g/ml~\cite{yellowbook}.

The Acrylic Sphere is surrounded by a spherical tank with 20\,m radius containing ultrapure water and the PMT array mounted on the {\it Stainless Steel Lattice Shell} (SSLS). Around 5,000 of the large 20''-PMTs are dynode PMTs while the larger part of around 13,000 are PMTs with a microchannel plate instead of a dynode structure~\cite{Wang_2017}.
To reach the goal of a highly precise energy resolution of 3\% at 1\,MeV, the PMT array has a geometrical coverage of about 78\%.
Moreover, each large PMT is required to have a high and homogeneous photon detection efficiency at a level of 28\%.
This leads to a total light collection of about 1200\,photoelectrons (p.e.) at 1\,MeV kinetic electron energy.
Scintillation photons from events with energies of a few MeV are detected usually in the single p.e. regime of each PMT. This minimizes non-linearity effects in the PMT readout electronics.

The steel sphere of the CD is contained in a {\it Water Pool}, which is equipped with around 2,400 20''-PMTs, and serves as a muon veto.
In addition to that, a {\it Top Tracker} (TT) made up of plastic scintillator strips, covers the water cylinder partially on top to measure the incoming direction of muons. 

\section{Modelling of the non-linear liquid scintillator light response}
\label{sec:model}

\subsection{Liquid scintillator non-linearity}
\label{sec:model_nl}
\subsubsection{Ionization quenching of scintillation light}
\label{sec:quench}
 The non-linearity of the scintillation light yield with energy deposit is known as ionization quenching. Empirically, the light yield per unit depth $\D X$ is expressed by the Birks' formula~\cite{Birks}:
\begin{equation}
\label{eq:bk}
  \left<\frac{\D L}{\D X}\right>=L_0 \frac{ \left<\frac{\D E}{\D X}\right>}{1+kB\left<\frac{\D E}{\D X}\right>},
\end{equation}
where $L$ is the light yield, $L_0$ is the scintillation light yield normalization, $\left<\D E/\D X\right>$ is the average energy loss of the particle per unit depth, and $kB$ is the Birks' material constant~\cite{Birks}. The parameter $L_0$ describes therefore the scintillation light yield under the absence of the quenching effect. The overall amount of the scintillation light $\left< L(E_\text{kin})\right>$ from the deposition of the kinetic energy $E_{\text{kin}}$ of particle in the LS can then be expressed as
\begin{equation}
\label{eq:pathint}
\left<L(E_\text{kin}) \right>=\int_\text{path} \left<\frac{\D L}{\D X}\right> \D X.
\end{equation}
The latter equation can be then reformulated using $\D X = \left<\D E/\D X\right>^{-1} \D E$ as
\begin{equation}
\label{eq:bkint}
  \left<L(E_\text{kin})\right> = L_0 \int_{0}^{E_\text{kin}}\frac{ 1}{1+kB\left<\frac{\D E}{\D X}\right>\left(E^\prime\right)} \D E^\prime.
\end{equation}
In the following, the contribution to the {\it visible energy} $E_\text{vis}$ from scintillation light will be determined from this equation. The {\it non-linearity} of the energy scale will be then defined as the dependence of the ratio $E_\text{vis}/E_\text{dep}$ on the particle's kinetic energy $E_\text{kin}$.
In the case of electrons and gammas, the particle's deposited energy $E_\text{dep}$ is equal to its kinetic energy.

To describe the suppression of the light emission in the LS, one needs to know the energy dependent energy loss per unit depth in the scintillation medium. For electrons, it is described by the M{\o}ller model, while the Bhabha model describes the energy loss of positrons~\cite{PDG}. The full formulas for these models can be found in Appendix~\ref{app:dedx}.

To apply the energy loss equations for the JUNO scintillator, one needs to find appropriate values for the density correction $\delta(E)$, the mean excitation energy $I$, and for the ratio of the atomic number to the atomic mass number $Z/A$. The LS volume is dominated by LAB, so the following calculations assume LAB as the only component of the scintillator, having
a chemical composition $\text{C}_{17}\text{H}_{29}$.
The parameter $Z/A$ is therefore approximated by $\left<Z/A\right> = \left<Z\right>/\left<A\right>$ by taking the atom-abundancy weighted mean of the LAB-molecule, as it is suggested in \cite{PDG}. The value is calculated to be $\left<Z/A\right> = 0.087\,\frac{\text{mol}}{\text{g}}$. The mean excitation energy $I = 58.9\,\text{eV}$ is taken from the output of the \textsc{EStar}-tool~\cite{estar}.

\begin{figure}[htb]
  \centering
  \includegraphics[width=0.95\textwidth]{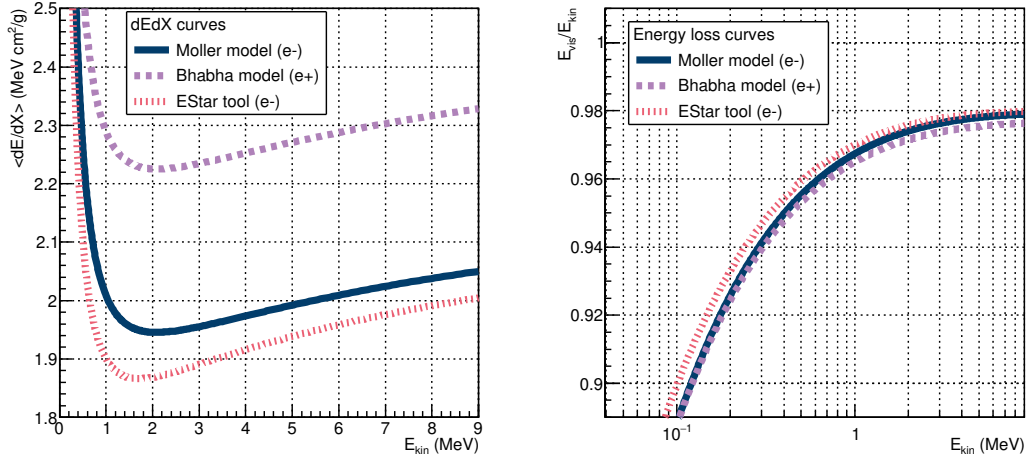}
  \caption{Comparison between the three different energy loss models (M{\o}ller, Bhabha, and \textsc{EStar}-tool) as a function of the kinetic energy. Left: The energy loss per unit depth $\left<\D E/\D X\right>$. Right: The respective non-linearity curves $E_\text{vis}/E_\text{kin}$, obtained via integration of the Birks' formula (Eq.~\ref{eq:bkint}) with $kB = 0.01$\,cm$^2$/MeV/g and a light yield normalization of $L_0=1$.}
  \label{fig:dedx_quench}
\end{figure}

The left plot in Fig.~\ref{fig:dedx_quench} shows a comparison between the energy losses $\left<\D E/\D X\right>$ calculated using the M{\o}ller and the Bhabha models, and evaluated via the \textsc{EStar}-tool. The same density correction term was applied in all three cases based on the evaluation in~\cite{Sternheimer}. Then, using the Eq.~\ref{eq:bkint}, one can obtain the respective $E_\text{vis}/E_\text{kin} = f(E_\text{kin})$ non-linearity curves.  The integration was evaluated numerically under the usage of the Gauss-Legendre method implemented in the ROOT framework~\cite{ROOT}.
A typical value of $k_B = 0.01$\,cm$^2$/MeV/g was used here and will be used in the later evaluations in this paper. The exact value for the JUNO scintillator was not conclusively measured so far. For the better visibility of the quenching impact, the scintillation light yield normalization is set to $L_0=1$.
With this normalization, the visible energy from scintillation light only is equal to the deposited energy under the assumption of no quenching.

The resulting three non-linearity curves, shown in the right plot of Fig.~\ref{fig:dedx_quench}, are similar for all three different $\left<\D E/\D X\right>$ energy loss models. We note, that the validity of the chosen quenching model needs to be evaluated based on the future calibration data.

 \subsubsection{Cherenkov light}
\label{sec:Cher}
Additionally to scintillation light, also Cherenkov light is produced by charged particles in the scintillation medium.
As it is usually created at small photon wavelengths~\cite{franktamm}, it is mostly absorbed and re-emitted by the scintillation medium~\cite{DYBLS}.
Due to this conversion, it can not be separated easily from the scintillation light and must be included in the non-linearity model.
The amount of Cherenkov light depends on the kinetic energy and the mass of the particle.
Therefore, the same amount of Cherenkov light is assumed for electrons and positrons.

\begin{figure}[t]
	\includegraphics[width=0.95\textwidth]{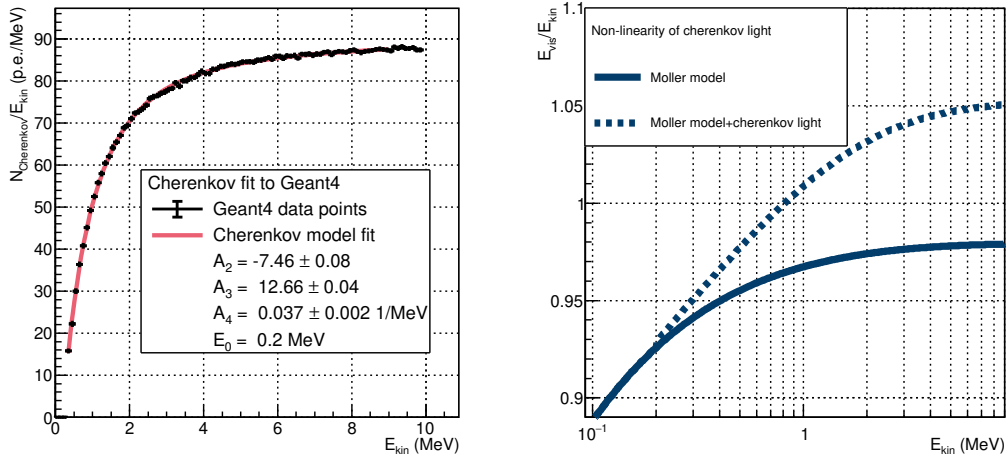}
  \caption{Left: Number of detected Cherenkov photons per $E_{\text{kin}}$ as a function of kinetic energy based on the JUNO Geant4 simulation of electrons (black points). The solid red line shows the fit of Eq.~\ref{eq:oleg} to the Monte Carlo data points. Right: Effect of the Cherenkov contribution to the LS non-linearity $E_\text{vis}/E_\text{kin}$, as a function of kinetic energy, using the M{\o}ller model for the ionisation loss calculation. The solid line shows the case with only the scintillation light, while the dashed line the case including also the Cherenkov light.
}
	\label{fig:cherenfit}
\end{figure}

To model the amount of detected Cherenkov photons in the detector, a simple empirical expression is applied, which was used~\cite{olegBx} in the solar-neutrino analysis of the Borexino experiment using Pseudocumene-based LS:
\begin{equation}
\label{eq:oleg}
\frac{N_{\text{Cherenkov}}(E_{\text{kin}})}{E_\text{kin}}=\left(\sum_{n=1}^3 A_n x^{n}\right)\cdot\left(\frac{1}{E_\text{kin}}+A_{4}\right)
\end{equation}
with 
\begin{equation*}
\label{eq:olegx}
  x = \ln\left(1+\frac{E_{\text{kin}}}{E_0}\right).
\end{equation*}
To evaluate the model of detected Cherenkov light and to estimate the parameters $A_i$, Eq.~\ref{eq:oleg} is fitted to the detected number of Cherenkov photons based on the JUNO Geant4 simulation~\cite{Lin_2017} for electrons.
As the parameters of Eq.~\ref{eq:oleg} were found to be highly correlated in the fit, the parameters $A_0$ and $A_1$ were fixed to $0$. The best fit values for the parameters $A_i$ with $i\in\{2,3,4\}$, which are shown with the fit on Fig.~\ref{fig:cherenfit}, are used in the following analysis.
The Cherenkov threshold was also estimated from the simulation output to be $E_0$ = 0.2\,MeV.

While the amount of Cherenkov light can be fully determined from Eq.~\ref{eq:oleg} after the determination of its parameters, the normalization of scintillation light and therefore the relative contribution of Cherenkov light still needs to be determined.
This is also taken from the JUNO Geant4 simulation. A simulation of electrons without quenching at 1\,MeV yields a ratio of
\begin{equation*}
 \label{eq:CherScintRatio}
  \frac{L_{\text{Cherenkov}}}{L_{\text{Scintillation}}}(1\,\text{MeV}) =\frac{50.39\pm0.07}{1217.6\pm0.2}= (4.142\pm0.006)\%,
\end{equation*}
with a pure statistical uncertainty originating in the amount of simulated data.
This number is used here to fix the relative contribution of Cherenkov light to the visible energy at 1\,MeV kinetic energy.
Due to this choice of normalization, the non-linearity ratio $E_\text{vis}/E_\text{kin}$ can be larger than $1$, if Cherenkov light is included in the following.

The JUNO Geant4 simulation is not assumed here to yield the amount of Cherenkov light with a high precision.
However, it is used here to obtain parameter values of Eq.~\ref{eq:oleg} and a normalization.
These values can be seen illustrative here, as they need to be evaluated in future calibration studies.

\subsection{Algorithmic calculation of the gamma non-linearity}
\label{sec:model_gammasim}

\subsubsection{Algorithm for calculating secondary electron energies}
\label{sec:secondary_el}

To develop a non-linearity model for positrons based on the electron non-linearity model, one needs to combine the non-linearity model of the positron itself and the non-linearity model of the two annihilation gammas.
The annihilation gammas interact with electrons of the scintillator with several different processes creating secondary electrons. To describe the energy loss of gammas, an algorithm was developed to obtain the energies of the secondary electrons, which are responsible for the production of scintillation light.
In the regime of gamma energies below $E_\gamma=2\cdot m_ec^2$, these are the photoelectric effect, Rayleigh-scattering, and Compton-scattering~\cite{PDG}.
Above $E_\gamma=2\cdot m_ec^2$, the gamma can also produce an electron-positron pair in an interaction with a nucleus or an electron.

\begin{figure}[t]
  \centering
    \includegraphics[width=0.7\textwidth]{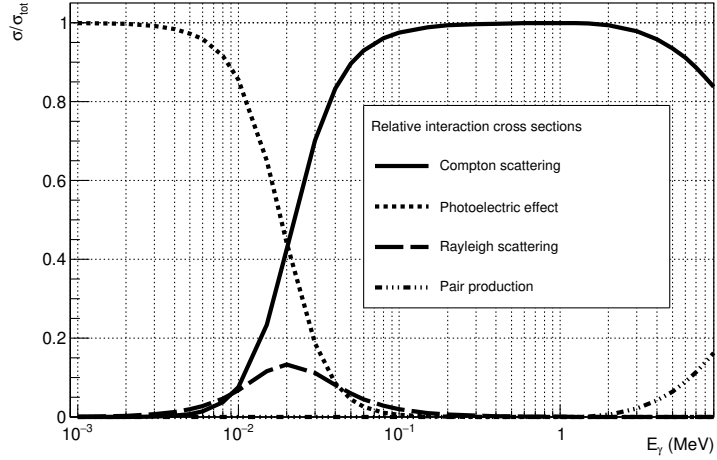}
    \caption{Relative cross sections of gamma scattering processes in LAB from~\cite{xcom}.}
	  \label{fig:gammaxsec}
\end{figure}

To simulate the behaviour of the initial gamma, the scattering process is determined first.
This was done by using the relative contribution of the process to the total interaction cross section as the probability to undergo that process.
The relative cross sections are shown in Fig.~\ref{fig:gammaxsec}.
This provides the probabilities of the gamma to interact with the respective process in the medium.
For higher energies above $E=2\cdot m_ec^2$, pure Compton-scattering was assumed as the cross section for electron-positron pair production is small compared to the cross section for Compton scattering for typical energies of a few MeV in JUNO~\cite{xcom}.

In the case of Rayleigh-scattering, the gamma does not loose energy and is just re-emitted in a different direction.
As the simulation does not consider the spatial behaviour of the gammas but only the energy, the process of Rayleigh scattering is not considered.

In the case of the photoelectric effect, the full energy of the gamma is transferred to the electron of the scintillation medium.
The gamma gets absorbed and the algorithm is stopped, if the photoelectric effect occurs.

\begin{figure}[!htb]
  \centering
    \includegraphics[width=0.7\textwidth]{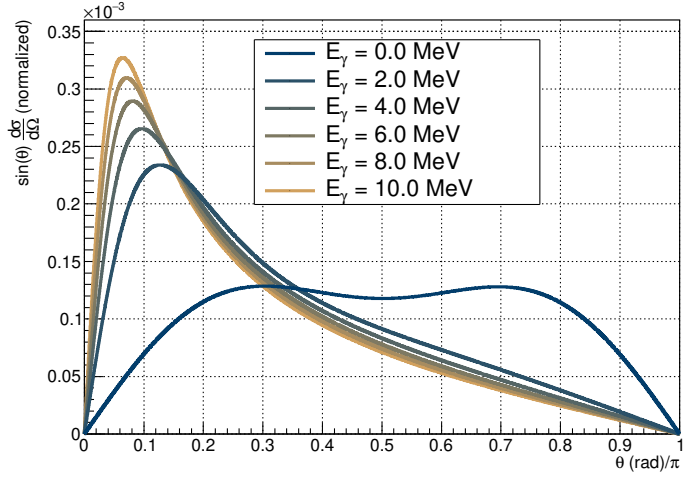}
    \caption{The evaluation of the Klein-Nishina cross section Eq.~\ref{eq:kleinnishina}. As it is used as a probability distribution, it is normalized to 1 and weighted with $\sin(\theta)$ as the integration is evaluated over $\D \cos{\theta}$.}
	  \label{fig:kleinnishina}
\end{figure}

Compton-scattering is the dominant process for initial gammas with an energy above $\approx$20\,keV in LAB.
For the process of Compton-scattering on an electron at rest, the distribution of scattering angles is calculated first.
The distribution of scattering angles is determined via the Klein-Nishina differential cross section~\cite{KleinNishina}:
\begin{equation}
  \frac{\mathrm{d}\sigma}{\mathrm{d}\Omega} =\frac{\mathrm{d}\sigma}{\mathrm{d}\phi \,\mathrm{d}\cos\theta} = \frac{\alpha^2}{2m_e^2}\left(\frac{E^\prime_\gamma}{E_\gamma}\right)\left[\frac{E^\prime_\gamma}{E_\gamma}+\frac{E_\gamma}{E^\prime_\gamma}-\sin^2\theta\right].
  \label{eq:kleinnishina}
\end{equation}
In this formula, $\alpha \approx 1/137$ is the electromagnetic fine structure constant, $E_\gamma$ is the gamma energy before scattering, $E^\prime_\gamma$ is the gamma energy after scattering, and $\theta$ is the scattering angle.
The energy loss ratio follows
\begin{equation}
  \frac{E^\prime_\gamma}{E_\gamma} = \frac{1}{1+\frac{E_\gamma}{m_e}(1-\cos\theta)}.
  \label{eq:kleinnishinaenergy}
\end{equation}
Since the total cross section in an scattering angle interval is proportional to the scattering probability, the normalized Klein-Nishina differential cross section is the probability density function (PDF) for the angular distribution shown in Fig.~\ref{fig:kleinnishina}.

The energy of the gamma after scattering is obtained from Eq.~\ref{eq:kleinnishinaenergy} by inserting a random angle following the angular distribution shown in Fig.~\ref{fig:kleinnishina}.
The kinetic energy of the scattered electron follows energy conservation:
\begin{equation}
  E^\prime_\text{e,kin} = E_\gamma - E^\prime_\gamma.
  \label{eq:knelectronenergy}
\end{equation}
After each Compton scattering the calculation is repeated until the gamma is absorbed via the photoelectric effect or has an energy of less than $E_{\text{min}}=250$\,eV.
This minimal energy was chosen to be the same as the default one used by the Geant4-software for electromagnetic processes~\cite{Geant4}.

\subsubsection{Validation and cross check}
\label{sec:gammasim_val}

\begin{figure}[!htb]
  \centering
  \includegraphics[width=\textwidth]{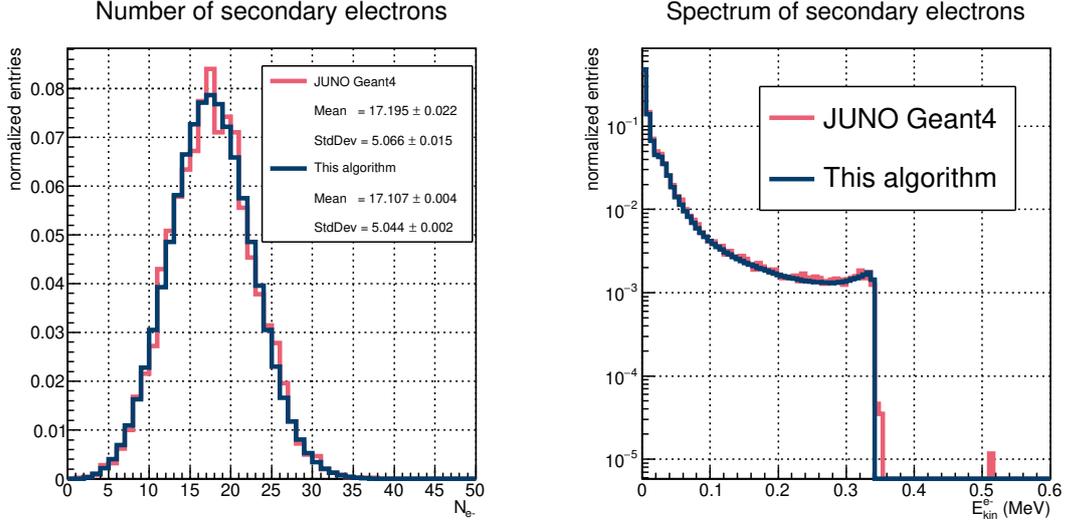}
  \caption{Comparison of the presented algorithm for secondary electrons (Sec.~\ref{sec:secondary_el}, solid blue line) with the JUNO Geant4 simulation~\cite{Lin_2017} (solid red line) for $E_{\gamma}$ = 511\,keV. The left side shows the number of secondary electrons and the right side shows their overall energy spectrum.}
  \label{fig:validation}
\end{figure}

The algorithm for calculating the distributions and energies of secondary electrons, as it was presented in the previous Section, is validated against the JUNO Geant4 simulation. We have chosen gammas with $E=m_ec^2=511$\,keV energy simulated in the detector center, in order to avoid border effects at the acrylic vessel. Figure~\ref{fig:validation} 
compares the distribution of the number of secondary electrons and their full energy spectrum for the case of our algorithm (solid blue line) and the JUNO Geant4 simulation (solid red line).
The complete physics list of the JUNO Geant simulation can be found in Appendix~\ref{app:sniper}.
For a better comparison, the histograms obtained with our algorithm were scaled down to match the amount of data in the Geant4 simulation. In general, the number of secondary electrons is not a reliable number for comparison, as it can depend on the so called \textit{production cuts}, {\it i.e.} the lower energy cut, at which the tracking of the mother particle is stopped~\cite{Geant4user}.
As the lower energy limit here is given by the dominance of the photoelectric absorption at low energies, these production cuts are not reached and the number of secondary electrons is a reliable quantity for comparison.
As we can see in the left part of 
Fig.~\ref{fig:validation}, the distribution of the number of secondary electrons shows a reasonable agreement with only slight differences of the mean and the standard deviation.
The right part of 
Fig.~\ref{fig:validation} demonstrates the excellent agreement of the overall energy spectrum of the secondary electrons.
This comparison approves the reliability of the presented algorithm.

\subsubsection{Computational benefits}
\label{sec:benfits}

\begin{figure}[t]
  \centering
    \includegraphics[width=0.7\textwidth]{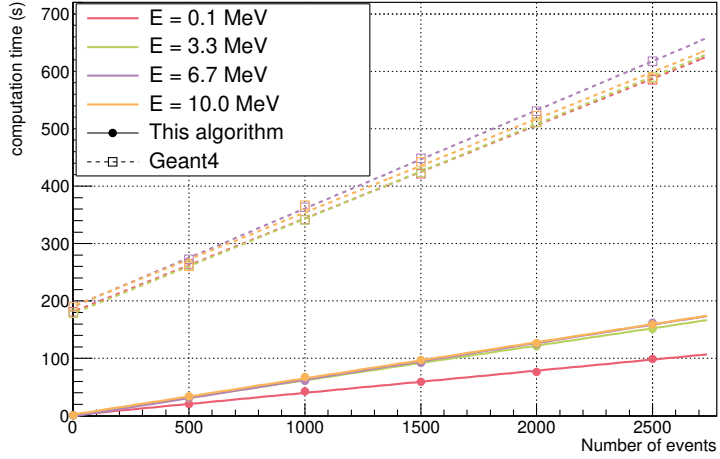}
    \caption{Comparison of the presented algorithm for secondary electrons (Sec.~\ref{sec:secondary_el}, solid lines) with the JUNO Geant4 simulation~\cite{Lin_2017} (dashed lines) for $E_{\gamma}$ = 0.1, 3.3, 6.7, and 10.0\,MeV in terms of the computational time. The lines represent estimations for the parameters $T_0$ and $T_\text{event}$ from Eq.~\ref{eq:cpu_time} for each $E_{\gamma}$ energy, obtained by a fit with equal weights for each data point. One can clearly see a large gain on the CPU time using our algorithm, due to differences both in the start-up time $T_0$ as well as in the $T_\text{event}$ time needed per event.}
	  \label{fig:cpu}
\end{figure}

In the process of calculating the NL from gammas or positrons (see Sec.~\ref{sec:positron}) using the results of this article, the production of a representative sample of secondary electrons, dominates the computing time.
To estimate the gain in computation time of this algorithm with respect to the evaluation by Geant4, both algorithms were run for different initial gamma energies and number of generated events.
All of these runs were executed on the same machine by an AMD Opteron\textsuperscript{TM} Processor 6238. In the JUNO Geant4-framework, the computation time is usually dominated by the propagation of optical photons, when the default settings are used. The production of these optical photons has been disabled. For the computation time one would expect, that each algorithm needs a certain amount of start-up time $T_0$ for its setup before each event needs about the same time $T_\text{event}$ to be processed.
The total computation time is expected to follow 
\begin{equation}
\label{eq:cpu_time}
T_\text{comp}=T_0 + T_\text{event}\cdot N,
\end{equation}
with $N$ being the number of processed events.
For both algorithms, the computation times for the energies of 0.1\,MeV, 3.3\,MeV, 6.7\,MeV, and 10.0\,MeV, as well as for the event numbers of 1, 500, 1000, 1500, 2000, and 2500 events were evaluated.
The parameters $T_0$ and $T_\text{event}$ were estimated then for each energy separately. The results are then  summarized graphically in Fig.~\ref{fig:cpu} and numerically in Table~\ref{tab:cpu}. The large gain in computation time is directly visible.
As described in~\cite{Lin_2017}, the JUNO Geant4 framework contains a comprehensive description of the detector geometry and the physics processes, which are relevant in the JUNO experiment.
It serves the purpose of being a general tool for simulating events in a broad energy range to study analysis methods and particle interactions with a precise model of the detection effects in the JUNO detector.
The higher complexity, caused by the higher universality, results especially in a higher start-up time, but as well in a higher time needed per event. 
Extrapolating the values from Table~\ref{tab:cpu} one would need around 4.5\,h to simulate 100\,000 gammas at 0.1\,MeV using the JUNO Geant4 simulation and only about 1\,h using the presented algorithm.

Apart from the gain in computational time, the presented algorithm has benefits in easier maintenance.
While the JUNO Geant4 framework is a complex framework built upon a software stack of external programs, the presented algorithm consists of single C++ class, which uses methods from the ROOT framework~\cite{ROOT}.

\begin{table}[ht]
  \centering
  \caption{Comparison of the presented algorithm for secondary electrons (Sec.~\ref{sec:secondary_el}) with the JUNO Geant4 simulation~\cite{Lin_2017} for $E_{\gamma}$ = 0.1, 3.3, 6.7, and 10.0\,MeV in terms of the computational time. The estimated values for the start-up time $T_0$ and the computational time needed per event $T_\text{event}$ as estimated from Fig.~\ref{fig:cpu}.}
  \label{tab:cpu}
    \begin{tabular}{ c | c | c | c | c }
      &   \multicolumn{2}{c|}{This algorithm}                                 &  \multicolumn{2}{c}{JUNO Geant4}  \\[5pt] \hline
      Energy\,(MeV) &   $T_0$\,(s)                        & $T_\text{event}$\,(s)       &  $T_0$\,(s)                   & $T_\text{event}$\,(s) \\[5pt] \hline
      0.1           &   1.41                                 & 0.0386                           &  181.23                            & 0.1624 \\[5pt]
      3.3           &   1.38                                 & 0.0604                           &  178.02                            & 0.1650 \\[5pt]
      6.7           &   -1.10                                 & 0.0640                           &  189.77                            & 0.1712 \\[5pt]
      10.0          &   2.40                                 & 0.0629                           &  190.93                            & 0.1631 \\[5pt] 
\end{tabular}
\end{table}

\subsubsection{Results for the non-linearity model of gammas}
\label{sec:NL_gammas}

To evaluate the scintillator non-linearity model for gammas from the simulation of the secondary electron spectrum, the sum of all light emissions from secondary electrons was taken.
The non-linearity of the secondary electrons was evaluated using the Birks' law (Eq.~\ref{eq:bkint}) and the energy loss via the M{\o}ller model (Eq.~\ref{eq:moller}).
It was assumed that there are no correlated effects in the light production between different secondary electrons and each of them can be treated individually.

\begin{figure}[t]
  \centering
    \includegraphics[width=\textwidth]{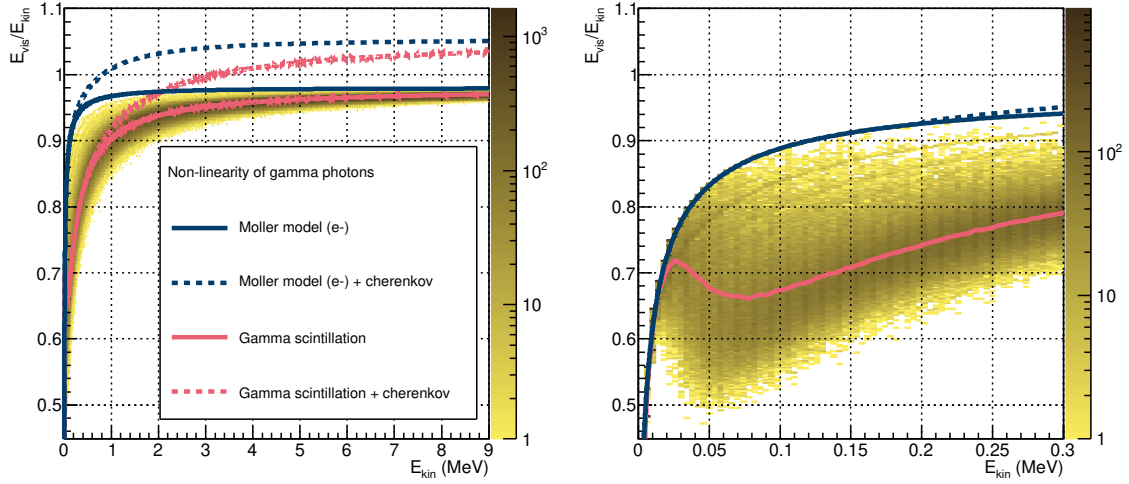}
    \caption{The results for the non-linearity $E_\text{vis}/E_\text{kin}$ model for gammas calculated with the presented algorithm. The distribution of the non-linearity of all gammas without the Cherenkov light is shown in yellow, while the corresponding average non-linearity is shown in solid red. The red dashed curve shows the average non-linearity of gammas with the Cherenov light included. For comparison, the electron non-linearity curve resulting from the M{\o}ller model is shown in blue, without (solid) and with (dashed) lines. The right plot shows the zoom of the left plot in the energy range below $E_\text{kin} < 0.3$\,MeV.}
	  \label{fig:nonlingamma}
\end{figure}

In Figure~\ref{fig:nonlingamma} one can see in yellow the evaluation of $E_\text{vis}/E_\text{kin}$ for about $1.5 \cdot 10^{7}$ gammas simulated without the Cherenkov light in the energy range  $E_\text{kin} = E_{\gamma}$ from 1\,keV to 9\,MeV. The right plot shows the zoom of the left plot in the energy range below $E_\text{kin} < 0.3$\,MeV. The average visible energy for the gammas without the Cherenkov light is shown with the solid red line, while the dashed red line represents the case with the Cherenkov light included. For comparison, the blue lines show the non-linearity model for electrons, again without (solid lines) and with (dashed lines) the Cherenkov light. For the same incident kinetic energy, the visible energy of electrons is always higher (or equal at very low energies) than the visible energy for gammas.
This is because the gammas only create visible energy due to secondary electrons,
having lower energies and thus, higher quenching (Fig.~\ref{fig:dedx_quench} right), with respect to the electron of the same kinetic energy as the original gamma. At very low energies below 20\,keV, the photoelectric effect becomes dominant, as it can be seen in Fig.~\ref{fig:gammaxsec}.
This causes the non-linearity curves for electrons and for gammas to be the same, as the full kinetic energy of the gamma is transferred to the electron. In the transition region at around 70\,keV, a local minimum of the nonlinearity can be seen in the right plot in Fig.~\ref{fig:nonlingamma}.

\subsection{Non-linearity model for positrons}
\label{sec:positron}
\subsubsection{Positron non-linearity at rest}
\label{sec:posi_atrest}

The energy deposition of positrons at energies of a few MeV happens usually in two steps. First, positrons deposit energy in the scintillator due to ionization and create scintillation light similar to the energy loss of electrons.
Additionally, a positron annihilates afterwards with an electron of the detector material to produce gammas.
In this section is is assumed, that positrons annihilate at rest after depositing their total kinetic energy due to ionization.
The resulting two gammas have therefore a total energy of $E_{\gamma} = m_ec^2 = 511$\,keV each.
To calculate the non-linearity curve for these positrons, one needs to combine the non-linearity curve resulting from the Bhabha model in Fig.~\ref{fig:dedx_quench} and the non-linearity of the two gammas at $E_\gamma=m_ec^2$:
\begin{equation}
E_{vis}^\text{e+}(E_\text{kin}) = E_\text{Bhabha}^{\text{ion}}(E_\text{kin}) + 2\cdot E_\text{vis}^{\text{Gamma}}(m_{e}c^2) .
\end{equation}

\begin{figure}[!htb]
  \centering
    \includegraphics[width=\textwidth]{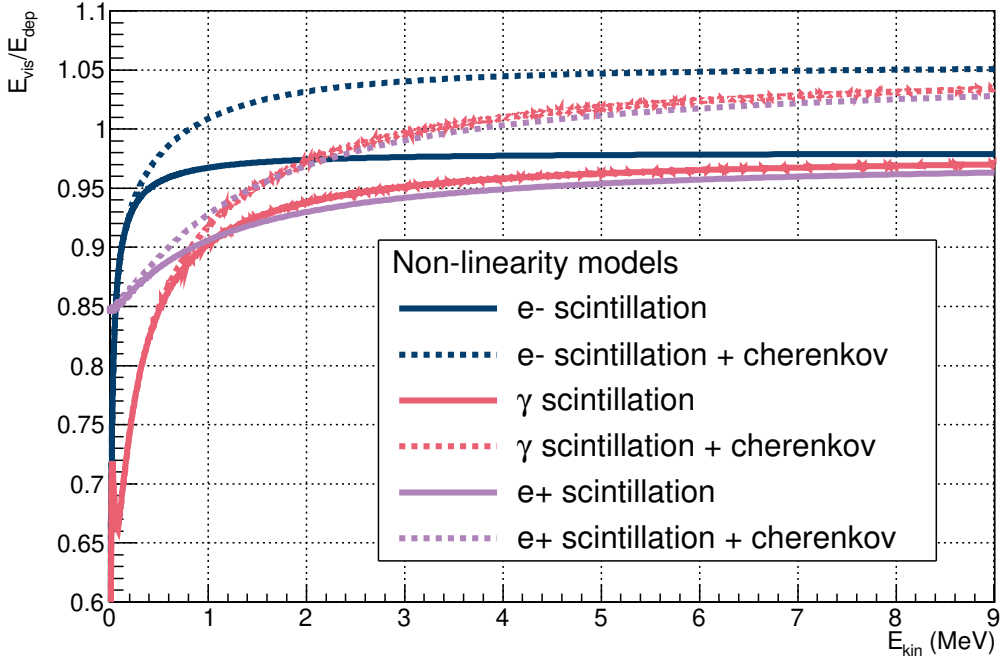}
    \caption{The non-linearity $E_{\text{vis}}/E_{\text{dep}}$ ($E_\text{dep}$ is the total deposited energy) models for positrons (purple), electrons (blue), and gammas (red). The solid curves show the non-linearity curves for scintillation light only, while the dashed curves include also the Cherenkov light.}
	  \label{fig:positron}
\end{figure}

The result of this evaluation together with the models for the electron non-linearity and the gamma non-linearity are shown in Fig.~\ref{fig:positron}.
Again, the dashed lines show the full deposited energy of scintillation light and Cherenkov light combined, while the solid lines represent the scintillation light only.
Here, the non-linearity is expressed as $E_{\text{vis}}/E_{\text{dep}}$, where
the variable $E_\text{dep}$ is used for the total deposited energy. For electrons and gammas it is $E_\text{dep} = E_\text{kin}$, while for positrons it is $E_\text{dep} = E_\text{kin}+2m_ec^2$.

\subsubsection{Evaluation of the full JUNO Geant4 simulation}
\label{sec:posi_higherorders}

\begin{figure}[!htb]
  \centering
    \includegraphics[width=\textwidth]{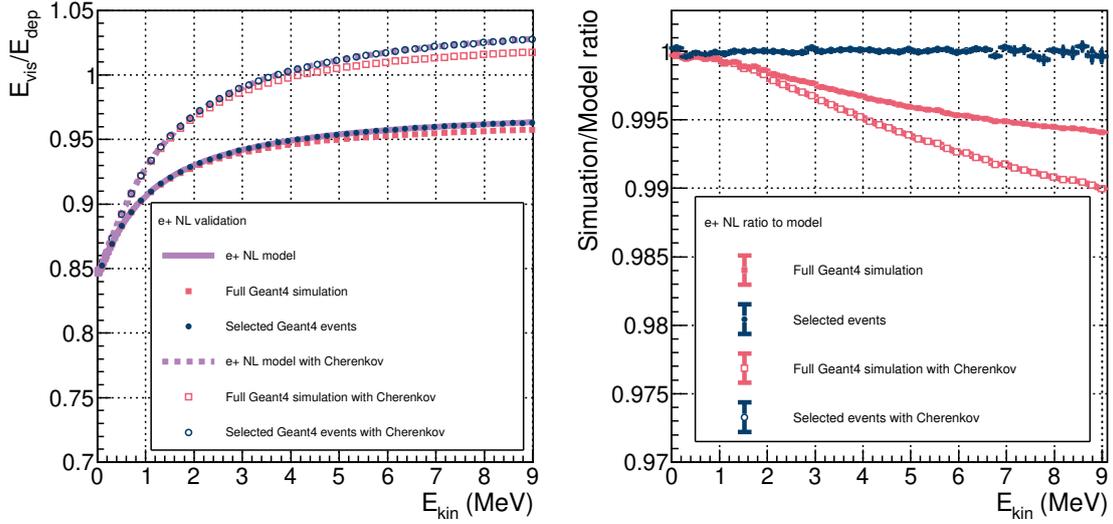}
    \caption{Comparison of the positron NL curve from Sec.~\ref{sec:posi_atrest} to the NL obtained by using the particle set created by the JUNO Geant4 simulation~\cite{Lin_2017} under usage of the M{\o}ller- and the Bhabha model. The graphs on the left show the comparison of the non-linearity model from Sec.~\ref{sec:posi_atrest} (purple) to all simulation events (red) and selected simulation events, which followed the assumptions of Sec.~\ref{sec:posi_atrest} (blue). In the right graphs the ratios of the NL evaluated using simulated particles by Geant4 to the NL curve from Sec.~\ref{sec:posi_atrest} are shown.}
	  \label{fig:positron_FullEasy}
\end{figure}

The calculation of the previous section assumes that positrons always annihilate at rest after loosing their kinetic energy completely in the scintillation medium via ionization. The annihilation in flight, the forming of positronium, and the creation of gammas via Bremsstrahlung are not considered. If positronium is formed, there is a chance to form para-positronium (p-Ps) or ortho-positronium (o-Ps).
If o-Ps is formed in vacuum, it can not decay into two gammas, as o-Ps has a total spin of 1 and therefore needs to decay into an odd number of photons.
However in matter, several effects cause o-Ps to decay into two gammas instead of three~\cite{oPsFranco, positronium}.
These are e.g. magnetic effects which cause a spin-flip or positron pick-off by surrounding electrons.
Moreover, the creation of electron-positron pairs by gammas was not considered in our model. 
To study the impact of these effects, the JUNO Geant4 simulation was used to generate a comprehensive set of particles and their energy depositions created by an initial positron.
Also here, the detectable light is entirely produced by positrons and electrons.
To evaluate the amount of visible energy, the amount of scintillation light was evaluated from the integration of Birks' formula (Eq.~\ref{eq:bkint}) and the amount of Cherenkov light was evaluated from Eq.~\ref{eq:oleg} for each positron and electron.
These contributions were summed up to the visible energy of the full event.
To cross-check the validity of the simple positron model from Sec.~\ref{sec:posi_atrest}, a sub-sample of these events was selected, which follows the assumptions of Sec.~\ref{sec:posi_atrest}.
These are the annihilation at rest into two gammas after the total kinetic energy is deposited due to ionization.
The production of Bremsstrahlung as well as the o-Ps decay into three gammas was not considered in Sec.~\ref{sec:posi_atrest}.
In total 970\,000 events were simulated in the JUNO Geant4 simulation, from which 78\,884 events were contained in the sub-sample following the assumptions of Sec.~\ref{sec:posi_atrest}.
The comparison of presented model from Sec.~\ref{sec:posi_atrest} with the full JUNO Geant4 simulation as well as the selected sub-sample can be seen in Fig.~\ref{fig:positron_FullEasy}.

One can clearly see the difference between the positron non-linearity curve and the full simulation, while the selected sub-sample shows no clear difference to the positron non-linearity curve.
This approves that the selected sub-sample is well described by the results of Sec.~\ref{sec:posi_atrest}, like it is expected from the validation results of Sec.~\ref{sec:gammasim_val}.
It can be further seen, that the deviation of the full simulation to the model barely exceeds 1\%, which is the requirement for the accuracy of the non-linearity model in JUNO. Nevertheless, it is expected to have additional sources of uncertainties as the limited range of calibration sources, as well as limited calibration data.
Therefore, the effects of the annihilation in flight and Bremsstrahlung should to be studied further and be included in the model.

\begin{figure}[!htb]
  \centering
    \includegraphics[width=0.7\textwidth]{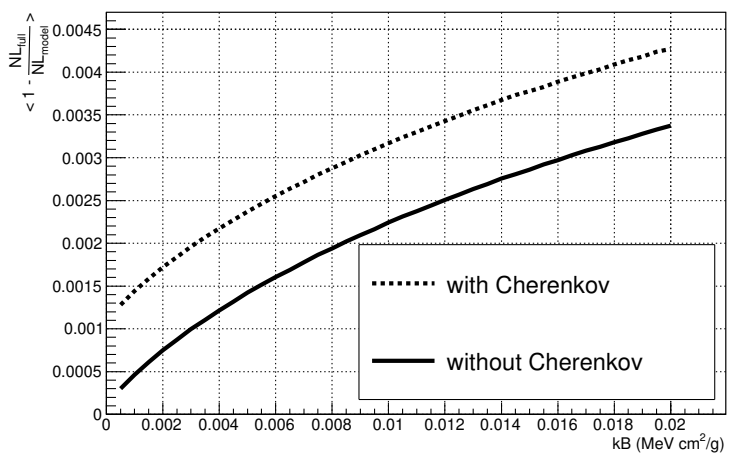}
    \caption{The deviation shown in Fig.~\ref{fig:positron_FullEasy} of the simple model to the JUNO Geant4 evaluation versus the Birks' constant averaged over the JUNO prompt spectrum under the assumption of normal neutrino mass ordering.}
	  \label{fig:diffVsKb}
\end{figure}

The averaged deviation of the Geant4 evaluation compared to the simple model in dependence of the Birks' constant is shown in Fig.~\ref{fig:diffVsKb}.
The average was taken over the expected JUNO prompt spectrum under the assumption of the normal neutrino mass ordering shown in Fig.~\ref{fig:junospectrum}.
One can see, that the average deviation is less than 0.5\% for the large evaluated range of $kB$.

\section{Summary and discussion}
\label{sec:summary}
As the accurate knowledge of the non-linearity is crucial for JUNO, it is of imminent importance to develop a good model for the positron non-linearity.
It was shown in this paper, that it is possible with easy methods to obtain a non-linearity model for electrons and gammas. For this an algorithm was shown to evaluate a precise spectrum of Compton electrons.

As only gamma sources are planned for the calibration of the JUNO experiment~\cite{yellowbook}, this evaluation can be used to determine the gamma non-linearity from the calibration data.
For this, the scintillation light normalization, the Birks's constant $kB$, as well as the Cherenkov curve parameters of Eq.~\ref{eq:oleg} need to be determined.
This can be used with the results of Sec.~\ref{sec:positron} to evaluate the positron non-linearity.

If the positron annihilates at rest without emitting Bremsstrahlung, the presented model shows good agreement with the presented Geant4 simulation.
However, if the positron annihilates in flight or produces Bremsstrahlung, there are clear differences of the model to the Geant4 simulation.
As the JUNO experiment has very stringent requirements of an accurate energy non-linearity description, these effects should be treated in further studies to be used in the later data-analysis.
Nevertheless, for sensitivity studies, the deviations are in a acceptable range to obtain a reasonable effect on the energy spectrum.

\section*{Acknowledgements}
This work was funded through the Recruitment Initiative by the Helmholtz Association of German Research Centers and through J\"ulich-OCPC Programme
for the Involvement of Postdocs in Bilateral Collaboration Projects with China. We thank Maxim Gonchar, Dmitry Naumov, and Konstantin Treskov from
Dzhelepov Laboratory of Nuclear Problems of JINR, Dubna for the fruitful discussions and the help in the reviewing process. We acknowledge the support by the JUNO collaboration for providing us their software framework.

\appendix
\section{Energy loss models}
\label{app:dedx}
%
For a quick reference, here are explicitly given the well-known energy loss equations from~\cite{PDG}.\\
The M{\o}ller model (electron-electron scattering):
\begin{multline}
  \label{eq:moller}
  \left<-\frac{dE}{dX}\right> = \frac{1}{2}K\frac{Z}{A}\frac{1}{\beta^2} \left[\ln\frac{m_ec^2\beta^2\gamma^2m_ec^2(\gamma-1)}{2I^2} +(1-\beta^2)-\frac{2\gamma-1}{\gamma^2}\ln2+\frac{1}{8}\left(\frac{\gamma-1}{\gamma}\right)^2-\delta(E)\right]
\end{multline}
The Bhabha model (positron-electron scattering):
\begin{multline}
  \label{eq:bhabha}
  \left<-\frac{dE}{dX}\right> = \frac{1}{2}K\frac{Z}{A}\frac{1}{\beta^2} \left[\ln\frac{m_ec^2\beta^2\gamma^2m_ec^2(\gamma-1)}{2I^2} +2\ln2-\frac{\beta^2}{12}\left(23+\frac{14}{\gamma+1}+\frac{10}{(\gamma+1)^2}+\frac{4}{(\gamma+1)^3}\right)-\delta(E)\right]
\end{multline}

\section{JUNO Geant4 Simulation: Physics List}
\label{app:sniper}
The JUNO Geant4 Simulation~\cite{Zou_2015}, which was used in this work, uses Geant4 in the version 9.4~\cite{Geant4, Geant4user}.
Due to the ongoing development in the JUNO collaboration, the used software does not represent the final simulation software used by the JUNO experiment.
For the simulation of electromagentic processes, the following physics list was used:

Electron:
\begin{itemize}
    \item G4eMultipleScattering
    \item G4LowEnergyIonisation
    \item G4LowEnergyBremsstrahlung
\end{itemize}

Gamma:
\begin{itemize}
    \item G4LowEnergyRayleigh
    \item G4LowEnergyPhotoElectric
    \item G4LowEnergyCompton
    \item G4LowEnergyGammaConversion
\end{itemize}

Positron:
\begin{itemize}
    \item G4eMultipleScattering
    \item G4eIonisation
    \item G4eBremsstrahlung
    \item G4PositroniumFormation
\end{itemize}

\printbibliography


\end{document}